\documentclass{article}
\usepackage{epsfig}
\usepackage{float}
\begin{document}
\title{Allowed Eta-Decay Modes and Chiral Symmetry}
\author{Barry R. Holstein$^a$\\[5mm]
Department of Physics\\
University of Massachusetts\\
Amherst, MA 01003\\and\\
Institut f\"{u}r Kernphysik\\
Forschungszentrum J\"{u}lich\\
D-52425 J\"{u}lich, Germany} 
\begin{titlepage}
\maketitle
\begin{abstract}

Recently, the development of chiral perturbation theory has allowed the
generation of rigorous low-energy theorems for various hadronic processes
based only on the chiral invariance of the underlying QCD Lagrangian.
Such techniques are highly developed and well-tested in the domain of 
pionic and kaonic reactions.  In this note we point out that with the
addition of a few additional and reasonable assumptions similar predictive
power is available for processes involving the eta meson.
\end{abstract}
\vfill
$^a$  email: holstein@physics.umass.edu
\end{titlepage}

\section{Introduction}
It has long been the holy grail for particle and nuclear knights 
to generate rigorous 
predictions from the Lagrangian of QCD
\begin{equation}
{\cal L}_{\rm QCD}=-{1\over 2}G_{\mu\nu}G^{\mu\nu}+\bar{q}(i\gamma_\mu
D^{\mu}-m)q\label{eq:qcd}
\end{equation}
where
\begin{eqnarray}
G_{\mu\nu}&=&\partial_\mu A_\nu -\partial_\nu A_\mu -ig[A_\mu ,A_\nu ]\nonumber\\
D_\mu q&=& (\partial_\mu -igA_\mu )q.
\end{eqnarray}
However, despite the ease with which one can write this equation, 
because of its inherent nonlinearity and the large value of the
coupling constant---$g^2/4\pi\sim 1$---progress in this regard has been
slow.  One approach---lattice gauge theory---holds great promise, but is 
currently
limited by the need for large computational 
facilities\cite{1}.  A second tack, that of
perturbative QCD, exploits the feature of asymptotic freedom---the vanishing of
the running coupling constant at high momentum 
transfer\cite{2}.  However, such predictions
are valid only for the very highest energy processes.  It is gratifying then to
see that in recent years a third procedure has become available, that of chiral
perturbation theory ($\chi$PT) which exploits the chiral symmetry of QCD and
allows rigorous predictive power in the case of low energy reactions.  This
technique, based on a suggestion due to Weinberg\cite{3}, was developed 
(at one loop level) during the last decade in an important series of papers by Gasser
and Leutwyler and others\cite{4}.  The idea is based on the feature that the QCD 
Lagrangian---Eq. \ref{eq:qcd}---possesses a global $SU(3)_L\times SU(3)_R$ (chiral) 
invariance in the limit of vanishing quark mass.  Such invariance is manifested
in the real world not in the conventional Wigner-Weyl fashion.
Rather, it is spontaneously
broken, resulting in eight light pseudoscalar 
Nambu-Goldstone bosons---$\pi, K,\eta$---which 
would be massless if the corresponding quark masses also vanished\cite{5}.
While the identification of this symmetry is apparent in 
terms of quark/gluon degrees of
freedom, it is not so simple to understand the implications of chiral invariance
in the arena of experimental meson/baryon interactions.

Early attempts in this direction were based on current algebra/PCAC 
methods\cite{6}, yielding relationships between processes differing in the
number of pions, {\it e.g.}
\begin{equation}
\lim_{q\rightarrow 0}<B\pi^a_q|{\cal O}|A>={-i\over F_\pi}<B|[F_5^a,{\cal O}]|A>
\end{equation}
where $F_\pi =92.4$ MeV is the pion decay constant\cite{7}.  However, it was
soon realized that the most succinct way to present these restrictions is in
terms of an effective chiral Lagrangian, the simplest (two-derivative) form of
which is, in the Goldstone sector\cite{8},
\begin{equation}
{\cal L}_{\rm eff}^{(2)}={\bar{F}^2\over 4}{\rm Tr}D_\mu UD^\mu U^\dagger 
+{\bar{F}^2\over 4}{\rm Tr}2B_0m(U+U^\dagger )+\cdots
\end{equation}
where
\begin{equation}
U=\exp\left({i\over \bar{F}}\sum_{j=1}^8\lambda_j\phi_j\right)
\end{equation}
is a nonlinear function of the pseudoscalar fields, $m=(m_u,m_d,m_s)_{\rm diag}$
is the quark mass matrix, 
\begin{equation}
2B_0={2m_K^2\over m_u+m_s}={2m_\pi^2\over m_u+m_d}
\end{equation}
is a phenomenological constant, $D_\mu=\partial_\mu-i[V_\mu,
]-i\{A_\mu, \}$ is the covariant derivative coupling to external
fields $V_\mu,A_\mu$, and
$\bar{F}$ is the pion decay constant in the limit of chiral symmetry.  Although
these are only two of an infinite number of terms, already at this level 
there exists 
predictive power--{\it e.g.}, tree level evaluation of ${\cal L}^{(2)}$
yields the familiar Weinberg predictions (at ${\cal O}(p^2,m^2)$) for 
S-wave $\pi -\pi$ scattering lengths\cite{9}
\begin{equation}
a_0^0={7m_\pi\over 32\pi F_\pi^2}\qquad a_0^2=-{m_\pi\over 16\pi F_\pi^2}
\end{equation}
which are approximately borne out experimentally.  Loop diagrams, of course,
produce terms of ${\cal O}(p^4,p^2m^2,m^4)$ and contain divergences.  However,
just as in QED such infinities can be absorbed into renormalizing phenomenological 
chiral couplings, and the most general "four-derivative" Lagrangian has been
given by Gasser and Leutwyler\cite{3}
\begin{eqnarray}
{\cal L}^{(4)}_{\rm eff}&=&L_1({\rm Tr}D_\mu UD^\mu U^\dagger )^2
+L_2({\rm Tr} D_\mu UD_\nu U^\dagger )^2\nonumber\\
& &+L_3{\rm Tr}(D_\mu UD^\mu U^\dagger)^2 +
L_4{\rm Tr}D_\mu UD^\mu U^\dagger ){\rm Tr}(m(U+U^\dagger ))\nonumber\\
& &+L_5{\rm Tr}D_\mu UD^\mu U^\dagger m(U+U^\dagger )
+L_6({\rm Tr}m(U+U^\dagger ))^2\nonumber\\
& &+L_7({\rm Tr}m(U-U^\dagger ))^2+
L_8 {\rm Tr}(mUmU+mU^\dagger mU^\dagger )\nonumber\\
& &+iL_9({\rm Tr}F^L_{\mu\nu}D^\mu UD^\nu U^\dagger +{\rm Tr}F^R_{\mu\nu}
D^\mu UD^\nu U^\dagger )\nonumber\\
& &+L_{10}{\rm Tr}F^L_{\mu\nu}UF^{R\mu\nu}U^\dagger 
\end{eqnarray}
where $F^L_{\mu\nu}, F^R_{\mu\nu}$ are external field strength tensors defined
via
\begin{eqnarray}
F^{L,R}_{\mu\nu}&=&\partial_\mu F^{L,R}_\nu -\partial_\nu F^{L,R}_\mu
-[F^{L,R}_\mu ,F^{L,R}_\nu ]\nonumber\\
F^L_\mu &=&V_\mu +A_\mu \qquad F^R_\mu =V_\mu-A_\mu .
\end{eqnarray}
Here the bare $L_i$ coefficients are themselves unphysical and are related 
to empirical quantities $L_i^r (\mu )$ measured at scale $\mu$ via
\begin{equation}
L_i^r(\mu )=L_i+{\Gamma_i\over 32\pi^2}\left({1\over \epsilon}+{\rm ln}
{4\pi\over \mu^2}+1-\gamma\right) ,
\end{equation}
where $\Gamma_i$ are constants defined in ref. 4b and $\epsilon=4-d$ is the
usual parameter arising in dimensional regularization, with $d$ being 
the number of dimensions.  Gasser and Leutwyler
have obtained empirical values for the phenomenological constants $L_1^r(\mu),
\ldots L_{10}^r(\mu)$, values for some of which are given in Table 1.

\begin{table}
\begin{center}
\begin{tabular}{|lllll|}  \hline
$L_1^r$ & $L_2^r$ & $L_5^r$ & $L_9^r$ & $L_{10}^r$ \\
\hline
$0.71\pm 0.28$ & $2.01\pm 0.37$ & $2.7\pm 0.3$ & $7.7\pm 0.2$ & $-5.2\pm 0.3 $\\
\hline
\end{tabular}
\caption{Empirical values of Chiral Expansion Parameters($\times 
10^{-3})$ with $\mu =m_\eta$.}
\end{center}
\end{table}

At the four-derivative level it is also necessary to include the contribution
of the anomaly, which in the case of coupling to the photon field $A_\mu$ has
the form\cite{10}
\begin{eqnarray}
& &\Gamma_{\rm WZW}(U,A_\mu )=\Gamma_{\rm WZW}(U)\nonumber\\
&+&{N_c\over 48\pi^2}\epsilon^{\mu\nu\alpha\beta }\int d^4x\left[eA_\mu{\rm Tr}
(Q(R_\nu R_\alpha R_\beta +L_\nu L_\alpha L_\beta))\right. \nonumber\\
&-&\left.ie^2F_{\mu\nu}A_\alpha {\rm Tr}\left(Q^2(R_\beta +L_\beta)+{1\over 2}
(QU^\dagger QUR_\beta +QUQU^\dagger L_\beta )\right)\right]\label{eq:anom}
\end{eqnarray}
where $R_\mu \equiv (\partial_\mu U^\dagger)U, L_\mu\equiv 
U\partial_\mu U^\dagger$ and $\Gamma_{\rm WZW}$ is independent of the photon
field and will not be needed for our purposes.  A corresponding form involving coupling to
a general nonabelian gauge field can also be written, but is lengthy and
will not be given here\cite{11}.

In a series of recent papers it has been conclusively demonstrated that this
chiral effective action formalism provides a succinct and successful description
of low energy electroweak interactions of pions and kaons\cite{12}.  Specifically, the
reactions given in Table 2 are successfully described in terms of 
GL parameters $L_9(\mu),L_{10}(\mu)$.  Clearly, there are far more reactions than parameters, and
this overdetermination enables one to construct {\it required} relationships 
between experimental
quantities, the empirical validity of which constitutes a strong test of the 
chiral formalism and indeed thereby of QCD itself\cite{13}.  Such tests are
found to be well satisfied, with the possible exception of the 
relationship between the axial structure function in radiative 
pion beta decay---$h_A$---and the charged pion 
polarizability---$\alpha_E^{\pi^+}$\cite{14}.  
However, recent work indicates that this may not be a problem and in any case
a number of experimental efforts are underway to retest this critical 
stricture\cite{15}.

\begin{table}
\begin{center}
\begin{tabular}{|llll|}
\hline
{\bf Reaction} & {\bf Quantity} & {\bf Theory} & {\bf Experiment} \\
$\gamma\rightarrow\pi^+\pi^-$ & $<r_\pi^2>$({\rm fm}$^2$) & 0.44$^a$ &
$0.44\pm 0.02$ \\
$\gamma\rightarrow K^+K^-$ & $<r_K^2>$({\rm fm}$^2$) & 0.44 & $0.34\pm 0.05$ \\
$\pi^+\rightarrow \pi^+\nu_e\gamma $ & $h_V(m_\pi^{-1})$ & 0.027 & $0.029
\pm 0.017$\\
{} & $h_A/h_V$ & $0.46^a$ & $0.46\pm0.08$ \\
$K^+\rightarrow e^+\nu_e\gamma$ & $(h_V+h_A)(m_\pi^{-1})$ & 0.038 &
$0.043\pm 0.003$\\
$\pi^+\rightarrow e^+\nu_ee^+e^-$ & $r_A/h_V$& 2.6 & $2.3\pm 0.6$ \\
$\gamma\pi^+\rightarrow\gamma\pi^+$ & $(\alpha_E+\beta_M)(10^{-4}({\rm fm}^3)$ &
0 & $1.4\pm 3.1$ \\
{} & $\alpha_E(10^{-4}({\rm fm}^3)$ & 2.8 & $6.8\pm 1.4$\\
$\gamma\gamma\rightarrow\pi\pi$ & $\alpha_E(10^{-4}({\rm fm}^3)$ & 2.8 &
$2.2\pm 1.6$\\
$K\rightarrow\pi e^+\nu_e$ & $\xi=f_-(0)/f_+(0)$ & -0.13 & $-0.20\pm 0.08$\\
{} & $\lambda_+({\rm fm}^2)$ & 0.067 & $0.065\pm 0.005$\\
{} & $\lambda_0({\rm fm}^2)$ & 0.040 & $0.050\pm 0.012$ \\
\hline
\end{tabular}

\caption{Chiral predictions and data in the radiative complex of 
transitions.  The superscript a indicates that this parameter was used
as input and is {\it not} a predicted quantity.}
\end{center}
\end{table}

\medskip

In the pion and kaon arena then one finds strong evidence for the correctness
and utility of chiral perturbative techniques, and it is an obvious next step
to attempt to extend this success into the eta sector, which is the 
subject of this note.  This examination of the eta system is important both as
a theoretical exercise and because of the existence of 
high intensity sources of etas\cite{16}.
In the next section then we study the ability of the chirally inspired
methods to
make reliable predictions for eta decay processes.  We examine only the 
"allowed" modes---$\eta\rightarrow 2\gamma , 3\pi , 2\pi\gamma, 
\pi 2\gamma ,3\pi\gamma$, {\it i.e.} those modes which can occur
assuming only isospin violation or the anomaly, eschewing the temptation to 
analyze ``rare'' processes such as
$\eta\rightarrow 3\gamma$, as these have been 
well-discussed elsewhere\cite{17}.
Finally, we summarize our results in a concluding section III.  

\section{Eta Decay Processes}

Strictly from a 
kinematic perspective, inclusion of the $\eta(547)$ as part of the chiral 
formalism should not be a problem, as the eta and kaon are 
roughly degenerate in mass, and as mentioned above the kaon (and pion) 
predictions
obtained in this way are quite successful.  Rather the real challenge
involves mixing with 
$\eta '(958)$, which lies outside the simple chiral $SU(3)_L\times SU(3)_R$
framework.  To lowest order things are simple---in the chiral limit the 
pseudoscalar mass spectrum would consist of a massless octet of Goldstone 
bosons plus a massive $SU(3)$ singlet ($\eta_0$).  With the breaking of 
chiral invariance the octet pseudoscalar masses become nonzero and are
related, at first order in chiral symmetry breaking,
by the Gell-Mann-Okubo formula\cite{gmo}
\begin{equation}
m^2_{\eta_8}={4\over 3}m_K^2-{1\over 3}m_\pi^2\approx (0.57\,\,{\rm GeV})^2\label{eqn:a}
\end{equation}
where $\eta_8$ is the eighth member of the octet.  At this same order in
symmetry breaking the singlet $\eta_0$ will in general mix with $\eta_8$
producing the physical eigenstates $\eta,\eta '$ given by
\begin{eqnarray}
\eta &=&\cos\theta \eta_8-\sin\theta\eta_0\nonumber\\
\eta '&=&\sin\theta\eta_8+\cos\theta\eta_0.
\end{eqnarray}
The mixing angle $\theta$ can be determined via diagonalization of the mass
matrix (written in the $\eta_8,\eta_0$ basis)
\begin{equation}
m^2=\left( \begin{array}{cc}
   m_{\eta_8}^2 & m_{08}^2 \\
   m_{08}^2       & m_{\eta_0}^2 \\
\end{array} \right)
\end{equation}
Here $m_{\eta_8}^2 $ is given in Eq. \ref{eqn:a} while $m_{08}^2,m_{\eta_0}^2$,
and $\theta$ are unknown.  Diagonalizing and 
fitting these parameters with the two known masses yields the results
\begin{equation}
\theta = -9.4^\circ ,\quad m_{08}^2=-0.44m_K^2,\quad
m_{\eta_0}=0.95\,{\rm GeV}
\end{equation}
However, there is good reason not to trust this simple and lowest order analysis, since
higher order chiral symmetry breaking terms can generate important modifications.
For example, taking the leading log correction arising
from Figure 1, we find\cite{18}
\begin{eqnarray}
m_{\eta_8}^2&=&{4\over 3}m_K^2-{1\over 3}m_\pi^2-{2\over 3}{m_K^2\over (4\pi 
F_\pi )^2}{\rm ln}{m_K^2\over \mu^2}\nonumber\\
&\approx & (0.61\,\,{\rm GeV} )^2 \quad {\rm for} \quad \mu =
1\,\,{\rm GeV},
\end{eqnarray}
for which diagonalization of the mass matrix yields 
\begin{equation}
\theta \approx -19.5^\circ ,\quad m_{08}^2=-0.81m_K^2,\quad
m_{\eta_0}=-0.90\,{\rm GeV}\label{eqn:b} 
\end{equation}
suggesting a doubling of the mixing angle.
\begin{figure}[htb]
\begin{center}
\epsfig{file=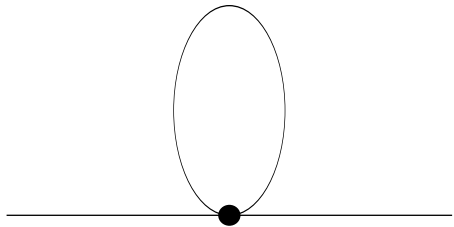,height=3cm,width=6cm}
\end{center}
\caption{Mass and wavefunction renormalization diagram.}
\end{figure}  
Of course, this is just an approximate result.  However, a full one loop
calculation using $\chi$PT yields essentially the same result\cite{4}, and 
consequently in our analysis below we shall use the value in Eq. \ref{eqn:b}, {\it i.e.}
\begin{equation}
\sin\theta \approx -{1\over 3}\qquad \cos\theta \approx {2\sqrt{2}\over 3}.
\end{equation}
It is also intriguing that this solution is consistent with the
assumptions of simple U(3) invariance wherein $\eta_8,\eta_0$ have the
same quark wavefunction, leading to
\begin{equation}
{m_{08}^2\over m_K^2}\simeq{2\sqrt{2}\over 3}\left(\hat{m}-m_s\over
\hat{m}+m_s\right)\simeq -0.9
\end{equation}

At this same (one-loop) level of symmetry breaking there is generated a shift
in the lowest order value of the pseudoscalar decay constant $F_P$.
Thus one finds from the diagrams in Figure 2, in
leading log approximation, 
\begin{eqnarray}
F_\pi &=&\bar{F}\left[1-{1\over 2}{m_K^2\over (4\pi F_\pi)^2}\ln {m_K^2\over 
\mu^2}\right] \approx 1.12\bar{F}\nonumber\\
F_{\eta_8}&=&\bar{F}\left[1-{3\over 2}{m_K^2\over (4\pi F_\pi )^2}\ln {m_K^2\over
\mu^2}\right] \approx 1.25 F_\pi \quad {\rm for} \quad \mu\approx
1\,\,{\rm GeV}.
\end{eqnarray}
Once again these estimates are in excellent agreement with those given in the full
one-loop analysis\cite{4}.  (We note in addition that the corresponding prediction
\begin{equation}
{F_K\over F_\pi}=1-{1\over 4}{m_K^2\over (4\pi F_\pi)^2}\ln {m_K^2\over \mu^2}
-{3\over 8}{m_\eta^2\over (4\pi F_\pi)^2}\ln {m_\eta^2\over \mu^2}\approx 1.22
\end{equation}
is quite consistent with the experimental value $1.22\pm 0.01$)
\begin{figure}[htb]
\begin{center}
\epsfig{file=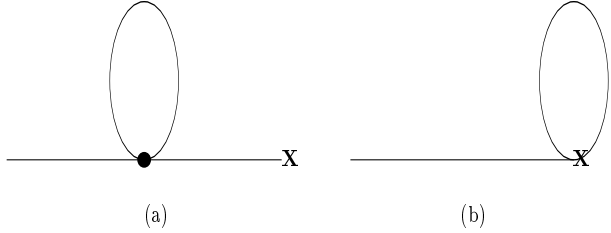,height=3cm,width=8cm}
\caption{Loop diagrams leading to renormalization of the pseudoscalar
decay constant.  Here the symbol $x$ indicates coupling to the axial current.}
\end{center}
\end{figure}

\subsection{\bf $\eta\rightarrow \gamma\gamma$}

With this introductory material in hand we can now confront the
subject of our report---that of eta decay.  First consider the dominant two-photon
decay mode, which to leading order arises due to the anomaly.  In the 
analogous $\pi^0\rightarrow \gamma\gamma$ case we find from Eq. \ref{eq:anom}
\begin{equation}
{\rm Amp}\equiv F_{\pi\gamma\gamma}(0)\epsilon^{\mu\nu\alpha\beta}\epsilon_\mu
k_\nu\epsilon'_\alpha k'_\beta \quad{\rm with}\quad F_{\pi\gamma\gamma}^{theo}(0)=
{N_c\alpha\over 3\pi F_\pi }=0.025\,\,{\rm GeV}^{-1}.\label{eq:theo}
\end{equation}
General theorems guarantee that this result is not altered in higher orders
of chiral symmetry breaking\cite{19} and, using the experimental value\cite{20}
\begin{equation}
\Gamma (\pi^0\rightarrow \gamma\gamma )= (7.7\pm 0.7)\,\,{\rm eV},
\end{equation}
we determine
\begin{equation}
F_{\pi\gamma\gamma}^{exp}=(0.0250\pm 0.0005)\,\,{\rm GeV}^{-1}
\end{equation}
in excellent agreement with the theoretically predicted value and eloquently
confirming the value $N_c=3$ as the number of colors.  Strictly speaking,
the prediction of the anomalous four-derivative Lagrangian Eq. \ref{eq:anom}
should be in terms of $\bar{F}$ rather than $F_\pi$.  Indeed the difference
between the value given in Eq. \ref{eq:theo} and the strict four-derivative prediction
involves terms of dimension six and is higher order in the chiral expansion.
Clearly, however, our prediction for $F_{\pi\gamma\gamma}(0)$, which arises from
what we shall term extended-$\chi$PT, is in excellent agreement with experiment.
Nevertheless, although very reasonable, this is {\it not} a firm prediction
of $\chi$PT itself.

The $\eta ,\eta '\rightarrow \gamma\gamma$ couplings 
also arise from the anomalous
component of the effective chiral Lagrangian and, in the extended $\chi$PT
approximation, have the values
\begin{eqnarray}
F_{\eta\gamma\gamma}(0)&=&{F_{\pi\gamma\gamma}(0)\over \sqrt{3}}
\left({F_\pi\over F_8}\cos\theta
-2\sqrt{2}{F_\pi\over F_0}\sin\theta\right)\nonumber\\
F_{\eta '\gamma\gamma}(0)&=&{F_{\pi\gamma\gamma}(0)\over \sqrt{3}}
\left( {F_\pi\over F_8}\sin\theta + 2\sqrt{2}{F_\pi\over F_0}\cos
\theta )\right) .
\end{eqnarray}
Using the experimental numbers\cite{fnt1}
\begin{eqnarray}
\Gamma (\eta\rightarrow\gamma\gamma )&=& (0.51 \pm 0.05)\,\,{\rm keV}  \qquad 
\Gamma (\eta'\rightarrow\gamma\gamma )= (4.7\pm 0.7)\,\,{\rm keV}
\end{eqnarray}
we find
\begin{equation}
\quad F_{\eta\gamma\gamma}(0)=0.0249\pm 0.0010\,\,{\rm GeV}^{-1} \qquad
F_{\eta'\gamma\gamma}(0)=0.0328\pm 0.0024\,\,{\rm  GeV}^{-1}\label{eq:data}
\end{equation}
In order to solve this system, we require an additional assumption
since there are three unknowns---$F_0,F_8,\theta$---but only two pieces
of data---Eq. \ref{eq:data}.  The standard approach at this point has been
to use the leading log prediction from one-loop chiral perturbation
theory---
\begin{equation}
{F_8\over F_\pi}=1-{m_K^2\over (4\pi F_\pi)^2}\ln {m_K^2\over
\mu^2}+{m_\pi^2\over (4\pi F_\pi)^2}\ln{m_\pi^2\over \mu^2}\simeq
1.30\quad {\rm at} \quad\mu\sim 1\, {\rm GeV}
\end{equation}
as input, and then to solve for $F_0,\theta$, yielding
\begin{equation}
\theta\simeq -20^\circ ,  \qquad {F_0\over F_\pi}\approx 1.04
\end{equation}
It is intriguing that these results from two phton decay are quite
compatable with those obtained from the one-loop analysis of the mass
matrix---{\it i.e.}, $\theta\simeq -20^\circ$ and $F_0/F_\pi$
consistent with the value of unity which one would obtain if the
singlet state and the pion were to have the same quark wavefunction.

Closely related to the above modes are the associated Dalitz 
decays---$\pi ,\eta ,\eta '\rightarrow \gamma e^+e^- $---which 
have been 
studied at DESY\cite{23}.  These are traditionally parameterized in terms
of a dipole slope parameter $\Lambda_P$ such that
\begin{equation}
{1\over \Gamma}{d\Gamma\over ds} = (1+ {s\over \Lambda_P^2})^2 
\quad {\rm with}\quad s=(p_+ +p_-)^2.
\end{equation}
The experimentally obtained values
\begin{equation}
\Lambda_\pi = 0.75\pm 0.03\,\,{\rm GeV}\qquad\Lambda_\eta=0.84\pm 0.06
\,\,{\rm GeV}\qquad 
\Lambda_{\eta'}=0.79\pm 0.04\,\,{\rm GeV}
\end{equation}
are in reasonable agreement with the vector dominance predictions\cite{24}
\begin{eqnarray}
\Lambda_\pi^2&=&m_{\rho ,\omega}^2\approx (0.77\,\,{\rm GeV})^2\nonumber\\
\Lambda_\eta^2&=&m_{\rho ,\omega}^2\left({3\cos\theta-
6\sqrt{2}\sin\theta\over
(5-2\xi^2)\cos\theta -(5+\xi^2)\sqrt{2}
\sin\theta}\right)\approx (0.75\,\,{\rm GeV})^2\nonumber\\
\Lambda_{\eta'}^2&=&m_{\rho ,\omega}^2\left( {3\sin\theta 
+6\sqrt{2}\cos\theta
\over (5-2\xi^2)\sin\theta 
+(5+\xi^2)\sqrt{2}\cos\theta}\right)\approx (0.83\,\,{\rm GeV})^2
\nonumber\\
& &{\rm where}\qquad \xi^2=(m_{\rho ,\omega}^2/m_\phi^2 )
\end{eqnarray}
These results are consistent with the observation that vector/axial dominance 
provides a remarkably successful representation of the values of the GL
parameters $L_i^r(\mu)$ obtained empirically\cite{25} and suggest an additional 
extension of our (already) extended $\chi$PT to include an effective 
vector-dominated Lagrangian\cite{26} 
\begin{equation}
{\cal L}_{\rm eff}={\cal L}_{\rm VPP}+{\cal L}_{V\gamma}+{\cal L}_{\rm VVP}
+{\cal L}_{PPP\gamma}
\end{equation}
with
\begin{eqnarray}
{\cal L}_{\rm VPP}&=&{-ig\over 4}{\rm Tr}V_\mu[\phi, \partial^\mu
\phi ]\nonumber\\
{\cal L}_{\rm V\gamma}&=&2eF_\pi^2gA^\mu\left(\rho_\mu +{1\over 3}\omega_\mu
-{\sqrt{2}\over 3}\phi_\mu \right)\nonumber\\
{\cal L}_{\rm VVP}&=&-{\sqrt{3}\over 4}g_{VVP}\epsilon^{\mu\nu\alpha\beta}
{\rm Tr}(\partial_\mu V_\nu\partial_\alpha V_\beta \phi )\nonumber\\
{\cal L}_{PPP\gamma }&=&{-ieN_c\over 24\pi^2F_\pi^3}
\epsilon^{\mu\nu\alpha\beta}A_\mu\partial_\nu\pi^+ \partial_\alpha\pi^-\label{eq:vdl}
\partial_\beta\pi^0
\end{eqnarray}
and 
\begin{equation}
g_{VVP}=-{3g^2\over 8\pi^2F_\pi},
\end{equation}
in order to understand the six-derivative component of the chiral expansion.
Here g is given by the KSRF relation as\cite{27}
\begin{equation}
g^2={m_\rho^2\over 2F_\pi^2}.
\end{equation}
In this formalism then the amplitude for $\pi^0\rightarrow \gamma\gamma $
is determined from the diagram in Figure 3 to be
\begin{equation}
F_{\pi\gamma\gamma}(0)={2e^2\over 3m_\omega^2m_\rho^2}(2eF_\pi g)^2
g_{\omega\rho\pi}={e^2\over 4\pi^2F_\pi}
\end{equation}
as the value required by the anomaly.  

With this background in hand, we
can now discuss a second important decay mode of the eta---that of 
$\eta\rightarrow \pi\pi\gamma$ which also arises from the anomalous
component of the Lagrangian.

\begin{figure}[htb]
\begin{center}
\epsfig{file=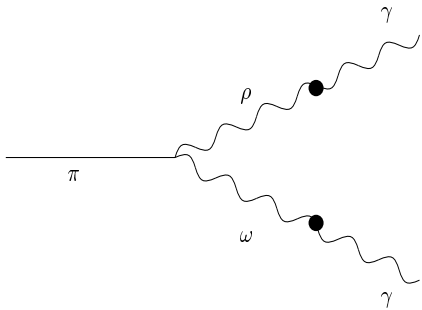,height=4cm,width=5.5cm}
\caption{Vector dominance diagram responsible for $\pi^0\rightarrow
\gamma\gamma$.}
\end{center}
\end{figure}

\subsection{\bf $\eta\rightarrow\pi^+\pi^-\gamma$}
In the previous section we performed an analysis of the QCD anomaly as
manifested in the two photon decay of the pseudoscalar mesons and, by
use of the one loop leading log value for $F_8/F_\pi$, were able to
determine a solution for the $\eta,\eta'$ mixing angle which is close
to that found in the mass analysis together with a value for $F_0/F_\pi$ which
is near that which results from the assumption that the $\eta_0$
and pion have the same wavefunction.  While this is somewhat
satisfying, it is intriguing to inquire whether one can assess the mixing
angle purely phenomenologically.  We shall show below that this
question can be answered in the affirmative, 
provided one utilizes the additional information available in
the anomalous decays $\eta,\eta'\rightarrow \pi^+\pi^-\gamma$. 
Such processes involving a photon coupled to three pseudoscalar mesons involve the
anomaly and, at zero four-momentum, are completely determined from 
Eq. \ref{eq:anom}.  However, inclusion of higher order effects generates structure and
study of such processes requires proper attention to issues of
unitarity and final state interactions.  Before considering
$\eta,\eta'$ decay, 
however, we consider first the closely related case of
$\gamma\rightarrow\pi^+\pi^-\pi^0$.  
At zero four-momentum the 
anomaly requires\cite{28}
\begin{eqnarray}
{\rm Amp}(3\pi -\gamma)&=&A(s_{+-},s_{+0},s_{-0})\epsilon^{\mu\nu\alpha\beta}
\epsilon_\mu p_{+\nu}p_{-\alpha}p_{0\beta}\nonumber\\
{\rm where}\quad A(0,0,0)&=&{eN_c\over 12 \pi^2F_\pi^3}=9.7\,\,{\rm GeV}^{-3}
\quad{\rm and}\quad s_{ij}=(p_i+p_j)^2\label{eq:vv}
\end{eqnarray}
and one might suspect that vector dominance could reproduce this result 
directly, as in the case of $\pi^0\rightarrow \gamma\gamma$ discussed above.
However, this turns out {\it not} to be the case.  Rather, use of the 
diagram shown in Figure 4a yields 
\begin{equation}
A(s,t,u)={eN_c\over 24\pi^2F_\pi^3}\left( {m_\rho^2\over m_\rho^2 -s}
+{m_\rho^2\over m_\rho^2-t}+{m_\rho^2\over m_\rho^2-u}\right)
\end{equation}
which at zero four-momentum is 50\% larger than the value given by the anomaly.
The resolution of this problem is well-known and arises from the a direct
$\gamma-3\pi$ coupling---Figure 4b---given in ${\cal L}_{PPP\gamma }$
whose origin presumably 
is from unspecified high-momentum-scale processes\cite{29}  .
Addition of this contribution to the
$\gamma -3\pi$ process yields an amplitude
\begin{equation}
A(s,t,u)={eN_c\over 12\pi^2F_\pi^3}\left[1+{1\over 2}\left({s\over m_\rho^2-s}
+{t\over m_\rho^2-t}+{u\over m_\rho^2-u}\right)\right]\label{eq:edep}
\end{equation}
which has the structure required by vector dominance, but also agrees with the
value required by the chiral anomaly at zero four-momentum.  Inclusion of 
loop corrections modifies Eq. \ref{eq:edep} slightly but does not 
change the general form.\cite{30}  

\begin{figure}[htb]
\begin{center}
\epsfig{file=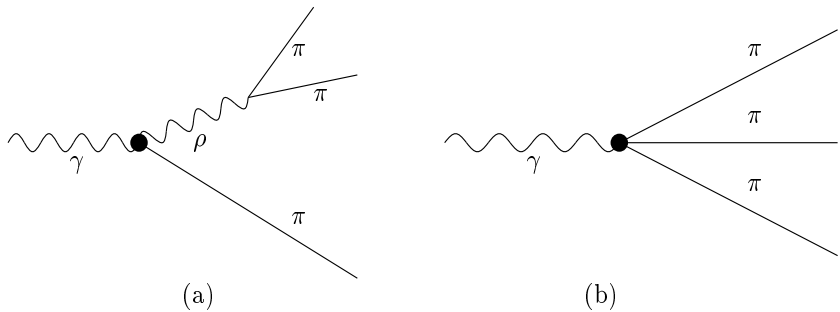,height=4cm,width=11cm}
\caption{Vector dominance diagrams responsible for the anomalous
process $\gamma\pi\rightarrow\pi\pi$.}
\end{center}
\end{figure}

The $\gamma -3\pi$ reaction has been studied experimentally via
pion pair production by the pion in the nuclear Coulomb field 
and yields a number\cite{31} 
\begin{equation}
A(0,0,0)_{\rm exp}=12.9\pm 0.9 \pm 0.5\,\,{\rm  
GeV}^{-3}
\end{equation}
in apparent disagreement with Eq. \ref{eq:vv} and suggesting the 
value $N_c\approx 4$!  This value was obtained, however, assuming no energy
dependence of the amplitude and is reduced to
\begin{equation}
A(0,0,0)_{\rm exp}=11.9\pm 0.9\pm 0.5\,\,{\rm GeV}^{-3}
\end{equation}
if Eq. \ref{eq:edep} is utilized, but is still too large.
The most likely conclusion is that this an experimental
problem associated with this difficult-to-measure process, but it has
recently been pointed out by Ametller, Knecht, and Talavera 
that an important electromagnetic
effect---the photon exchange diagram connecting $\pi^0\gamma\gamma^*$
and $\pi^+\pi^-\gamma$ vertices---can reduce this number by another
$1\times 10^{-3}$ or so\cite{akt}.  In any 
case a new high-precision experiment is clearly called for, and
this has been accomplished at JLab using the CLAS detector and the
reaction $\gamma p\rightarrow \pi^+\pi^- n$. 
Such a measurement has been also been proposed at Da$\Phi$ne\cite{32}.

The JLab experiment is currently being analyzed and this high
statistics measurement will require an equally careful theoretical
analysis in order to produce the desired extrapolation from the rho
resonance region, where most of the data has been obtained, to the
zero four-momentum point where the anomaly stricture obtains.  This
issue has been approached in a number of authors:
\begin{itemize} 
\item [i)] Holstein has used a simple matching of the one loop chiral 
correction to the rho dominance form\cite{hol}.  
\item [ii)] Truong has utilized a unitarization based upon the
Omnes-Muskhelishvili method\cite{tru}.
\item [iii)] Hannah unitarizes the amplitude using the inverse
amplitude procedure\cite{han}.
\end{itemize}
Regardless of the method used, the results are similar and are somewhat
robust.  The resulting value of the anomaly obtained 
in a preliminary analysis of the CLAS measurement are consistent, within
a significant uncertainty,
with the expected number---three.   However, a 
definitive value awaits further analysis.

Having warmed up on the $\gamma -3\pi$ process, it is now 
straightforward to construct the analogous $\eta\rightarrow\pi^+\pi^-\gamma$
amplitude.  Using the extended $\chi$PT assumption we find\cite{34}
\begin{equation}
{\rm Amp}(\eta\rightarrow\pi^+\pi^-\gamma )=B(s_{+-},s_{+\gamma},s_{-\gamma})
\epsilon^{\mu\nu\alpha\beta}\epsilon^*_\mu p_{+\nu}p_{-\alpha}k_{\gamma\beta}
\end{equation}
with the anomaly stricture yielding
\begin{equation}
B_\eta(0,0,0)={eN_c\over 12\sqrt{3}\pi^2F_\pi^3 }\left( {F_\pi\over F_8}\cos\theta
-\sqrt{2}{F_\pi \over F_0}\sin\theta \right)\label{eq:vda}
\end{equation}
However, the physical region for the decay---$4m_\pi^2\leq
s_{\pi\pi}\leq m_\eta^2$---is far from the zero-momentum point which
is constrained by the anomaly.  One indication of this fact is that
the decay rate obtained via neglect of momentum 
dependence---$\Gamma^{(0)}_{\eta\rightarrow\pi\pi\gamma}$=35.7 eV---is
significantly different from the experimental 
value---$\Gamma_{\eta\rightarrow\pi\pi\gamma}^{expt}=64\pm 6$ eV.  For
the $\eta'$ channel the situation is, of course, much worse.  The experimental
value---$\Gamma^{expt}_{\eta'\rightarrow\pi\pi\gamma}=61\pm 5$ keV---is a
factor of twenty larger than the 
value $\Gamma^{(0)}_{\eta'\rightarrow\pi^+\pi^-\gamma}$=3 KeV 
obtained via use of the simple anomaly prediction
\begin{equation}
B_{\eta'}(0,0,0)={eN_c\over 12\sqrt{3}\pi^2F_\pi^3 }\left( {F_\pi\over F_8}\sin\theta
+\sqrt{2}{F_\pi \over F_0}\cos\theta \right)\label{eq:vdb}
\end{equation}
Thus proper inclusion of momentum dependence is essential.
The $\eta\rightarrow\pi^+\pi^-\gamma$ spectrum was measured in the experiment
of Gormley et al.\cite{34} and was found to be approximately 
fit in terms of a pure (width-modified) $\rho$-dominated matrix element.
This result is {\it not} in agreement, however, with the simple vector 
dominance prediction---{\it cf.} Eq. \ref{eq:vdp}---which would require
\begin{equation}
|B(s,t,u)|_{\rm theo}^2\sim 1+3{s\over m_\rho^2}+\cdots
\end{equation}
and corresponds instead to 
\begin{equation}
|B(s,t,u)|^2_{\rm exp}\sim 1+2{s\over m_\rho^2}+\cdots .
\end{equation}
Thus a careful look at the unitarization procedure is called for.  

We begin by noting that a one-loop chiral perturbation
theory calculation gives
\begin{eqnarray}
B_\eta^{\rm 1-loop}(s,s_{\pi\pi})&=&B_\eta(0,0)[1+{1\over
32\pi^2F_\pi^2}
((-4m_\pi^2+{1\over3}s_{\pi\pi})\ln{m_\pi^2\over m_\rho^2}\nonumber\\
&+&{4\over
3}F(s_{\pi\pi})-{20\over 3}m_\pi^2+{3\over 2m_\rho^2}s_{\pi\pi}]
\end{eqnarray}
where
\begin{equation}
F(s)=\left\{
\begin{array}{ll}
(1-{s\over 4m_\pi^2})\sqrt{s-4m_\pi^2\over s}\ln{1+\sqrt{s-4m_\pi^2\over
s}\over -1+\sqrt{s-4m_\pi^2\over s}}-2 & s>4m_\pi^2 \\
2(1-{s\over 4m_\pi^2})\sqrt{4m_\pi^2-s\over s}\tan ^{-1}\sqrt{s\over
4m_\pi^2-s}-2& s \leq 4m_\pi^2
\end{array}\right.
\end{equation}
while the vector dominance picture ({\it cf.} Figure 5) yields
\begin{equation}
B_\eta(s_{\pi\pi})=B_\eta(0,0,0)\left(1+
{3\over 2}{s_{\pi\pi}\over m_\rho^2-s_{\pi\pi}}\right)\label{eq:vdp}
\end{equation}

\begin{figure}[htb]
\begin{center}
\epsfig{file=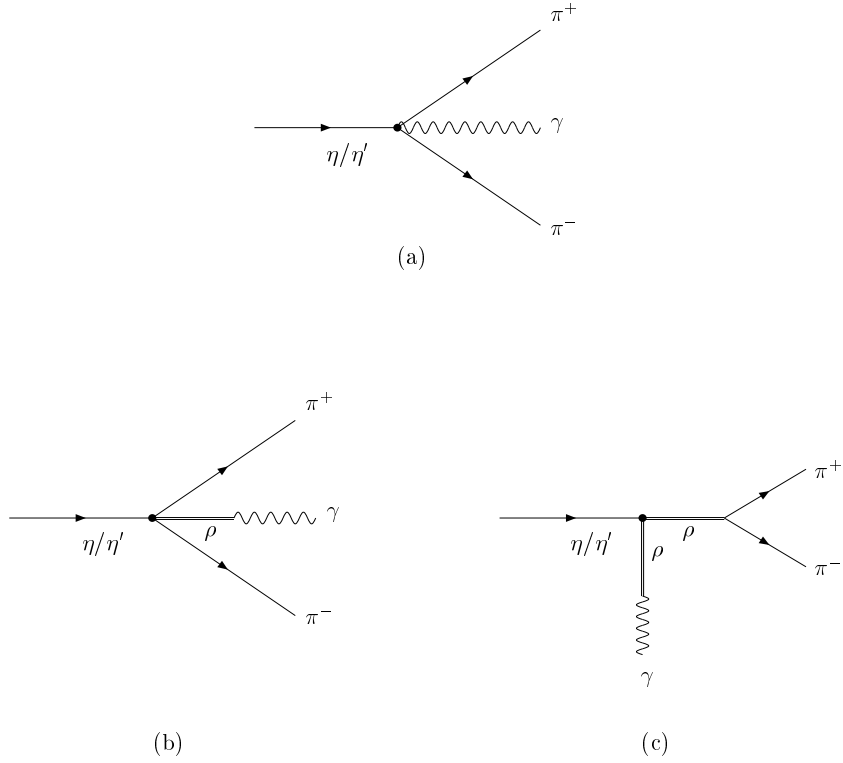,height=10cm,width=11cm}
\caption{Shown are contact (a) and VMD (b,c) contributions to
$\eta,\eta'\rightarrow\pi^+\pi^-\gamma$ decay.}
\end{center}
\label{fig1}
\end{figure}

Certainly, in order to treat the decay of the $\eta'$, one {\it must}
include unitarity effects via final state interactions. One very
obvious approach is simply to include the (energy-dependent) width of the 
rho-meson in the propagator in the vector-dominance form
Eq. \ref{eq:vda} via
\begin{equation}
{s_{\pi\pi}\over m_\rho^2-s_{\pi\pi}}\rightarrow
{s_{\pi\pi}\over m_\rho^2-s_{\pi\pi}-im_\rho\Gamma_\rho(s_{\pi\pi})}
\end{equation}
This use of vector width-modified vector dominance 
already makes an important difference from the simple anomaly---tree 
level---results (especially in the case of
the $\eta'$), changing the predicted decay widths from the values 
35 eV and 3 KeV quoted above to the much more realistic numbers
\begin{equation}
\Gamma^{theo-VM}_{\eta-\pi\pi\gamma}=62.3\,\,{\rm eV}, \qquad 
\Gamma^{theo-VM}_{\eta'-\pi\pi\gamma}=67.5\,\,{\rm KeV} 
\end{equation}
if the parameters
\begin{equation}
F_8/F_\pi=1.3,\qquad F_0/F_\pi=1.04,\qquad \theta=-20^\circ\label{eq:hh}
\end{equation} 
are employed.  However, this procedure
does not match onto the one-loop chiral form in the low energy limit.

In order to determine a form for the final state interactions which
matches onto both the one-loop chiral correction {\it and} to the vector
dominance result in the appropriate limits, we postulate an N/D
structure
\begin{equation}
B_{\eta-\pi\pi\gamma}(s,s_{\pi\pi})=B_{\eta-\pi\pi\gamma}(0,0)\left[
1-c+c{1+as_{\pi\pi}\over D_1(s_{\pi\pi})}\right]\label{eq:ii}
\end{equation}
where $D_1(s)$ is the Omnes function and is defined in terms of the
p-wave $\pi\pi$ phase shifts via\\cite{fnt2}
\begin{equation}
D_1(s)=\exp\left(-{s\over
\pi}\int_{4m_\pi^2}^\infty{ds'\delta(s')\over s'(s'-s-i\epsilon)}\right)
\end{equation}
and $a,c$ are free parameters to be determined.  In
order to reproduce the coefficient of the $F(s_{\pi\pi})$ function, which
contains the rho width, we require $c=1$.  On the other hand, matching
onto the VMD result at ${\cal O}(p^6)$ can be achieved by the choice 
$a=1/2m_\rho^2$.  Thus in the case of the $\eta$ the form is
completely determined.  Since the $\eta'$ spectrum is closely related
and is dominated by the presence of the rho we shall postulate an
identical form for the $\eta'$ case.  Using these forms we can
then calculate the decay widths assuming the theoretical values for
the anomaly.  Using the parameters given in Eq. \ref{eq:hh}
one finds, for example, 
\begin{eqnarray}
 i)D_1^{\rm exp}(s)  && \Gamma_{\eta-\pi\pi\gamma}=65.7\,\,{\rm eV}, \quad
\Gamma_{\eta'-\pi\pi\gamma}=66.2 KeV\nonumber\\
ii)D_1^{\rm anal}(s) && \Gamma_{\eta-\pi\pi\gamma}=69.7\,\,{\rm eV}, \quad
\Gamma_{\eta'-\pi\pi\gamma}=77.8 KeV
\end{eqnarray}
There is a tendency then for the numbers obtained via the analytic
form of the Omnes function to be somewhat too high.  

\begin{figure}[htb]
\leftline{
\begin{minipage}[t]{.47\linewidth}
\mbox{\leftline{\epsfig{file=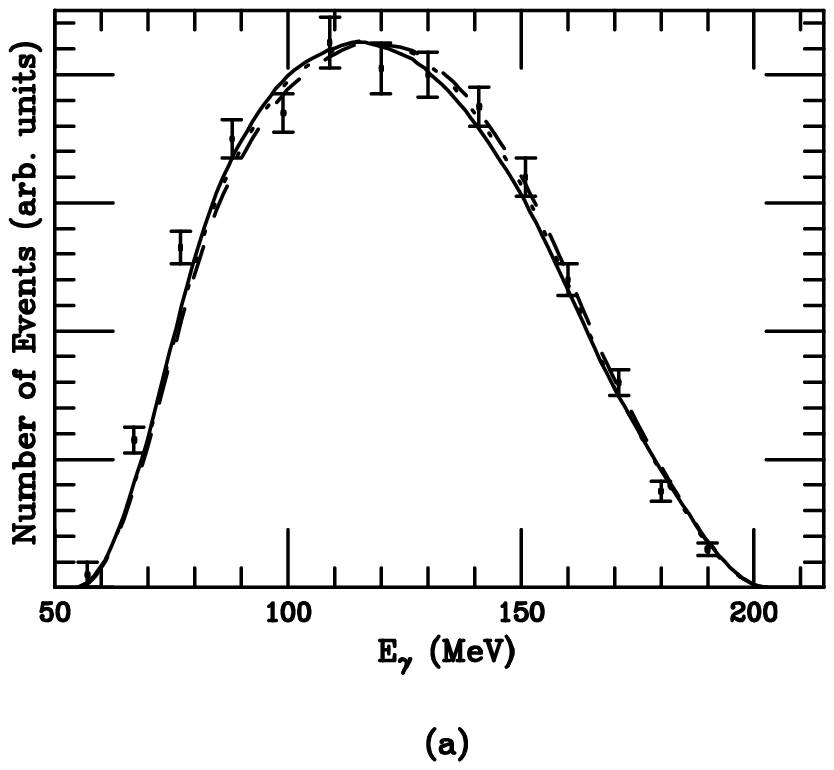,width=6.0cm}}
\hspace{0.6cm}
\rightline{\epsfig{file=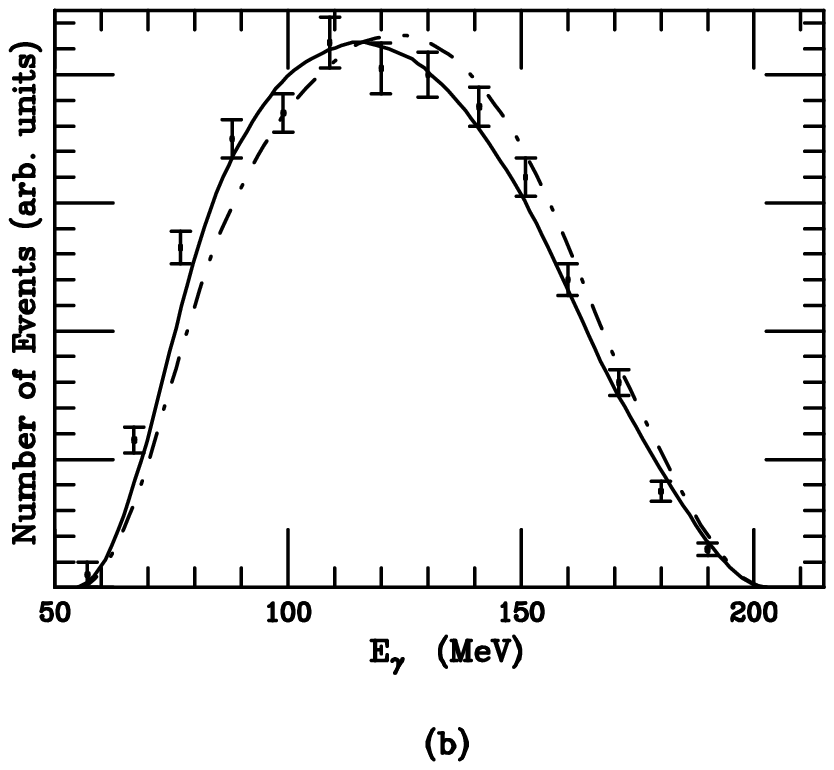,width=6.0cm}}}
\end{minipage}}
\caption{Shown is the photon spectrum in
$\eta\rightarrow\pi^+\pi^-\gamma$ from Gormley et al.\protect\cite{34} as well as
various theoretical fits.  In the Figure 6a, the dashed line represents the
(width-modified) VMD model.  The (hardly visible) dotted line and the
solid line represent the
final state interaction ansatz Eq.\ref{eq:ii} with use of the 
analytic and experimental version of the Omnes function respectively.
Figure 6b shows the experimental Omnes
function result (solid line) compared with the one-loop result (dotdash line).}
\label{fig2}
\end{figure}

\begin{figure}[htb]
\centering
\leavevmode
\centerline{\epsfbox{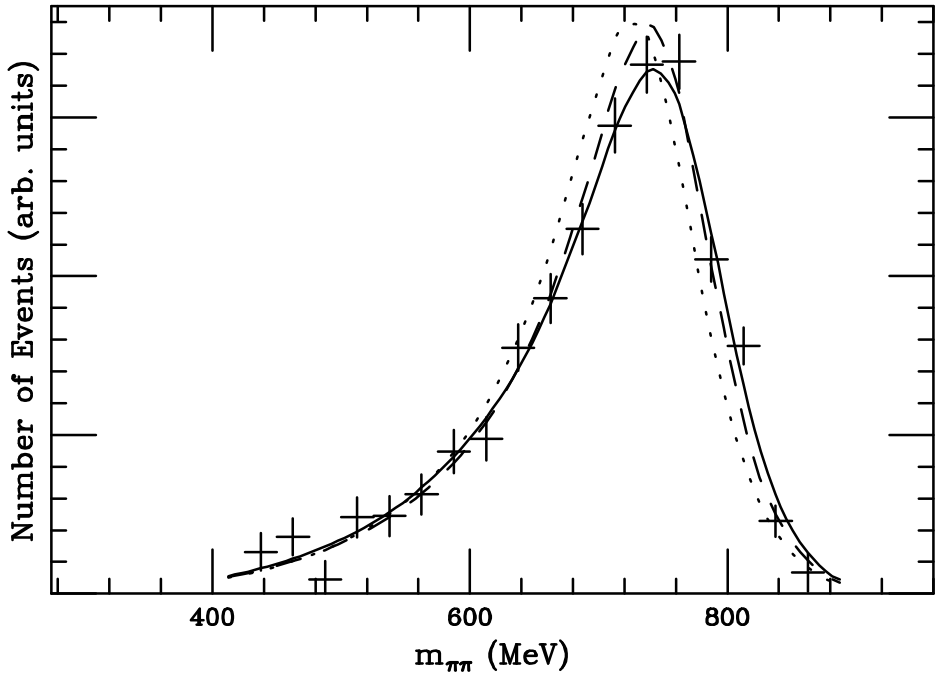}}
\caption{Shown is the photon spectrum in 
$\eta'\rightarrow\pi^+\pi^-\gamma$ from Abele et al.\protect\cite{83} as well
as various theoretical fits.  As in Figure 6a, the dashed line represents the
(width-modified) VMD model.  The dotted and solid lines represent 
the final state interaction ansatz Eq.\protect\ref{eq:ii}  with use 
of the analytic and experimental version of the Omnes function, respectively.
Here the curves have been normalized to the same number of events.}
\label{fig3}\end{figure}

We can also
compare the predicted spectra with the corresponding experimentally
determined values.  As shown in Figure 6, we observe that the
experimental spectra are well fit in the $\eta$ case 
in terms of both the N/D or the VMD
forms, but that the one-loop chiral expression does not provide an
adequate representation of the data.  In the case of the corresponding
$\eta'$ decay the results are shown in Figure 7, wherein we observe
that either the unitarized VMD or the use of $N/D_1^{\rm exp}$ provides
a reasonable fit to the data (we get $\chi^2$/dof=32/17 and 20/17,
respectively), while the use of the analytic form for the Omnes
function yields a predicted spectrum ($\chi^2$/dof=104/17) 
which is slightly too low on the high energy end.  However, for both 
$\eta$ and $\eta'$ we see that our simple ansatz---Eq.\ref{eq:ii}---provides 
a very satisfactory representation of the decay spectrum.
 Our conclusion in the previous section was that if the mixing angle and
pseudoscalar coupling constants were assigned values consistent with
present theoretical and experimental leanings, then the predicted
widths and spectra of
both $\eta,\eta\rightarrow\pi^+\pi^-\gamma$ are basically consistent with
experimental values.  Our goal in this section is to go the other way,
however.  That is, using the assumed N/D forms for the decay amplitude,
and treating the pseudoscalar decay constants $F_8,F_0$ as well as the
$\eta-\eta'$ mixing angle $\theta$ as free parameters, we wish to
inquire as to
how well they can be constrained purely from the experimental data
on $\eta,\eta'\rightarrow\gamma\gamma$ and
$\eta,\eta'\rightarrow\pi^+\pi^-\gamma$ decays, with reasonable
assumptions made about the final state interaction effects in these
two channels. 

On theoretical grounds, one is somewhat more confident about the
extraction of the threshold amplitude in the case of the lower energy
$\eta\rightarrow\pi^+\pi^-\gamma$ system.  Indeed, in this case the
physical region extends only slightly into the tail of the rho unlike 
the related $\eta'$ decay wherein the spectrum extends completely over the
resonance so that there exists considerable sensitivity to details 
of the shape.  Thus a first approach might be to utilize only the
two-photon decays together with the
$\eta\rightarrow\pi^+\pi^-\gamma$ width in order to determine the
three desired parameters.  In this fashion one finds the results shown
in Table 1.
\begin{table}
\begin{center}
\begin{tabular}{|c|c|c|c|} \hline
 &$F_8/F_\pi$    & $F_0/F_\pi$    & $\theta$\\ \hline
VMD & $1.28\pm 0.24$ & $1.07\pm 0.48$ & $-20.3^\circ\pm 9.0^\circ$\\
N/D$_1^{\rm anal}$ & $1.49\pm 0.29$ & $1.02\pm 0.42$ & 
$-22.6^\circ\pm 9.6^\circ$\\
N/D$_1^{\rm exp}$ & $1.37\pm 0.26$& $1.02\pm 0.45$& $-21.2^\circ\pm
9.3^\circ$\\ \hline
\end{tabular}
\caption{Values of the renormalized pseudoscalar coupling
constants and the $\eta-\eta'$ mixing angle using the
$\eta,\eta'-\gamma\gamma$ and $\eta-\pi\pi\gamma$ amplitudes in a
three parameter fit.}
\end{center}
\end{table}
We observe that the results are in agreement, both with each other and
with the chiral symmetry expectations---$F_8/F_\pi\sim 1.3$,
$F_0/F_\pi\sim 1$, and $\theta\sim -20^\circ$.  However, the
uncertainties obtained in this way are uncomfortably high.  

In order to ameliorate this problem, we have also done a maximum likelihood fit
including the $\eta'-\pi\pi\gamma$ decay rate, yielding the results
shown in Table 2. 
\begin{table}
\begin{center}
\begin{tabular}{|c|c|c|c|} \hline
 &$F_8/F_\pi$& $F_0/F_\pi$& $\theta$\\ \hline
VMD & $1.28\pm 0.20$& $1.07\pm 0.04$ & $-20.8^\circ\pm 3.2^\circ$\\
N/D$_1^{\rm anal}$& $1.48\pm 0.24$& $1.09\pm 0.03$& $-24.0^\circ\pm
3.0^\circ$\\
N/D$_1^{\rm exp}$&$1.38\pm 0.22$&$1.06\pm 0.03$&$-22.0^\circ\pm
3.3^\circ$\\ \hline
\end{tabular}
\caption{Values of the renormalized pseudoscalar coupling
constants and of the $\eta-\eta'$ mixing angle obtained from a maximum
likelihood analysis using the $\eta,\eta'-\gamma\gamma$ and
$\eta,\eta'-\pi\pi\gamma$ amplitudes.}
\end{center}
\end{table}
We observe that the central values stay fixed but that the error bars
are somewhat reduced.  The conclusions are the same,
however---substantial renormalization for $F_8\sim 1.3 F_\pi$, almost
none for $F_0\sim F_\pi$, and a mixing angle $\theta\sim -20^\circ$.
These numbers appear nearly invariant, regardless of the approach.

An interesting aside here is the recent observation by B\"{a}r and Wiese 
that the $\pi^0\rightarrow\gamma\gamma$ reaction alone does not verify
the three color hypothesis---in a careful analysis the $N_c$-dependence 
of the Wess-Zumino-Witten term is completely canceled by the 
$N_c$-dependent part of a Goldstone-Wilczek term, and that it is only the 
$\eta\rightarrow\pi^+\pi^-\gamma$ measurement which truly confirms the
result that $N_c=3$\cite{baw}.

Having above confirmed the basic correctness of the predictions of the anomaly
(and thereby of this important cornerstone of QCD) we move now to the important
three pion decay of the eta, which occurs independent of the anomaly and which
rather probes the conventional two- and four-derivative piece of the chiral
Lagrangian.

\subsection{\bf $\eta\rightarrow\pi\pi\pi$}
The decay of the isoscalar eta to the predominantly I=1 final state of the
three pion system occurs primarily on account of 
the d-u quark mass difference\cite{fnt3}, and the result arising from 
lowest order
chiral perturbation theory is well-known\cite{os}
\begin{equation}
{\rm Amp}(\eta_8\rightarrow\pi^a\pi^b\pi^c)=\delta^{ab}\delta^{c3}C(s_{ab},
s_{ac},s_{bc})+{\rm Permutations}
\end{equation}
where
\begin{equation}
C(s,t,u)=-{B_0(m_d-m_u)\over
3\sqrt{3}F_\pi^2}\left[1+{3(s-s_0)\over m_\eta^2-m_\pi^2}\right]\label{eq:qq}
\end{equation}
and we have defined 
\begin{equation}
s_{ab}=(p_a+p_b)^2\qquad{\rm and}\qquad s_0={1\over 3}(m_\eta^2+m_{\pi^+}^2
+m_{\pi^-}^2+m_{\pi^0}^2).
\end{equation}
Equivalently, we can write the prefactor of Eq. \ref{eq:qq} in a form
which respects the reparameterization invariance of Kaplan
and Manohar\cite{km}
\begin{equation}
{B_0(m_d-m_u)\over 3\sqrt{3}F_\pi^2}=-{1\over Q^2}{m_K^2\over m_\pi^2}(m_K^2-m_\pi^2)
\end{equation}
where
\begin{equation}
Q^2={m_s-\hat{m}\over m_d-m_u}{m_s+\hat{m}\over m_d+m_u}
\end{equation}
and $\hat{m}={1\over 2}(m_d+m_u)$ is the average $u,d$ quark mass.

Thus the decay $\eta\rightarrow 3\pi$ can be used in order to
determine the quantity $Q^2$.  Alternatively, if $Q^2$ is given from
some other process then the $\eta\rightarrow  3\pi$ amplitude is completely
determined.  The standard approach to such a determination is to utilize
the pseudoscalar meson masses via the relation
\begin{equation}
Q^2={m_K^2\over m_\pi^2}{m_K^2-m_\pi^2\over (m_{K^0}^2-m_{K^+}^2)_{\rm QCD}}
\end{equation}
where $(m_{K^0}^2-m_{K^+}^2)_{\rm QCD}$ is the nonelectromagnetic
component of the $K^0,K^+$ mass difference---
\begin{equation}
(m_{K^0}^2-m_{K^+}^2)_{\rm QCD}=(m_{K^0}^2-m_{K^+}^2)_{\rm expt}
-(m_{K^0}^2-m_{K^+}^2)_{\rm em}\label{eq:da}
\end{equation}
In order to evaluate the right hand side of Eq. \ref{eq:da} one
generally uses Dashen's theorem, which guarantees the identity of the
electromagnetic piece of the kaon and pion electromagnetic mass shifts
in the chiral symmetric limit\cite{da}
\begin{equation}
(m_{\pi^+}^2-m_{\pi^0}^2)=(m_{K^+}^2-m_{K^0}^2)_{\rm EM}.
\end{equation}
This simple assumption gives then
\begin{equation}
Q^2_{\rm Dashen}={m_K^2\over m_\pi^2}{m_K^2-m_\pi^2\over m_{K^0}^2-m_{K^+}^2+
m_{\pi^+}^2-m_{\pi^0}^2}=24.1
\end{equation}
and results in a prediction
\begin{equation}
\Gamma (\eta\rightarrow\pi^+\pi^-\pi^0)=66\,\,{\rm eV}
\end{equation}
in strong contradiction to the experimental result
\begin{equation}
\Gamma^{exp}(\eta\rightarrow\pi^+\pi^-\pi^0)=281\pm 28\,\,{\rm eV}.
\end{equation}
At first sight this would appear to be a rather strong and irreparable 
violation of a lowest order chiral prediction and therefore not salvagable
by the expected ${\cal O}(m_\eta^2/(4\pi F_\pi^2)^2)\sim 30\% $ corrections
from higher order effects.  However, this is not at all the case.  In fact
the one-loop and counterterm contributions were calculated by Gasser
and Leutwyler and were found to enhance the lowest order prediction by a
factor 2.6, yielding\cite{gle}
\begin{equation}
\Gamma^{theo}(\eta\rightarrow\pi^+\pi^-\pi^0 ) 
\approx 167\pm 50\,\,{\rm eV},\label{eq:kk}
\end{equation}
which is a significant improvement, but still somewhat too low.  The origin of
such a large correction lies primarily with $\eta_8,\eta_0$ mixing which
generates a factor $(\cos\theta -\sqrt{2}\sin\theta )^2\sim 2$ leaving
the expected $30\%$ corrections due to conventional higher order loop
and counterterm contributions.  However, recent work has indicated that
Eq. \ref{eq:kk} is probably a considerable underestimate 
due a significant violation
of Dashen's theorem.  Indeed, the Dashen requirement was derived in
the limit of chiral symmetry---$m_\pi^2=m_K^2=0$---and plausible estimates 
of chiral breaking effects have yielded the estimate\cite{dhw}
\begin{equation}
Q^2_{\chi -{\rm broken}}\approx 0.8Q^2_{\rm Dashen},\quad {\it i.e.}\quad
Q_{\chi -{\rm broken}}\sim 21.7\label{eq:pp}
\end{equation}
which corresponds to an additional $\sim$40\% enhancement of the chiral calculation
Eq. \ref{eq:kk}, {\it i.e.}
$\Gamma^{theo}(\eta\rightarrow\pi^+\pi^-\pi^0)
\sim 240\,\,{\rm eV}$,
and puts the result in the right ballpark.  There are at least two possible
sources for the existence of any remaining discrepancy.  One is the fact that
the estimate for the size of Dashen's theorem violation is just that---an 
estimate.  It is possible that the size of the violation is more significant than
that given in Eq. \ref{eq:pp}, leading to an even 
larger value for $\Gamma (\eta\rightarrow
\pi^+\pi^-\pi^0 )$.  A second possibility is that the simple one loop estimate
given in ref. 4 is not sufficient to include the full impact of final state 
interaction effects.  This has been demonstrated in other 
processes where the I=0 S-wave $\pi -\pi$ plays an important role, as it does
here\cite{tr}.  Indeed the closely related $K\rightarrow 3\pi$ reaction is one
such case\cite{ns}.  

In order to decide which---if either---possibility obtains it is necessary
to make careful spectral shape measurements in addition to simple
lifetime numbers.  Also it is necessary to confront such results with
precise theoretical calculations.  Phenomenologically, we expand the
decay amplitude about the center of the Dalitz plot as 
\begin{equation}
C(s,t,u)\equiv \alpha\left[1 +\beta y +\gamma y^2+ \delta
x^2+\ldots \right]
\end{equation}
where 
\begin{equation}
y={3(s-s_0)\over 2m_\eta \Delta_\eta}\qquad{\rm and}\qquad
x={\sqrt{3}(t-u)\over 2m_\eta \Delta_\eta}.
\end{equation}
where $\Delta_\eta=m_\eta-2m_{\pi^\pm}-m_\pi^0$ is the Q-value.
These parameters have been determined experimentally 
to be
\begin{eqnarray}
\mbox{Layter et al.}\cite{exp1}: \beta &=& 0.54\pm 0.007\qquad\gamma 
= 0.017\pm 0.014\qquad\delta =0.023\pm 0.016\nonumber\\
\mbox{Gormley et al.}\cite{exp2}: \beta&=& 0.585\pm 0.010\qquad \gamma
= 0.105\pm 0.015\qquad\delta
=0.03\pm 0.02\nonumber\\
\mbox{Amsler et al.}\cite{exp3}: \beta&=&0.470\pm
0.075\qquad\gamma=0.055\pm 0.135
\end{eqnarray}
to be compared to the one-loop chiral prediction 
\begin{equation}
\beta = 0.665 \qquad\gamma = 0.21 \qquad\delta = 0.04
\end{equation}
Clearly there is general (though certainly not excellent) agreement,
suggesting the importance of higher order scattering contributions.

These have been examined by two Swiss collaborations using dispersion
relation treatments in order to address the problem of higher
order three-body scattering effects.  The calculation of Anisovich and
Leutwyler quotes only the integrated decay rate which is is agreement
with experiment if the value $Q=22.7\pm 0.8$ is chosen\cite{al}---consistent
with the Dashen theorem violation calculated in \cite{dhw}.  Similarly
the integrated decay rate found in the Khuri-Treiman calculation by
Kambor, Wiesendanger, and Wyler agrees with the experimental rate if
the value $Q=22.4\pm 0.9$ is used\cite{kww}.  However, these authors
also quote values for the spectral shape
\begin{eqnarray}  
\beta&=&0.58\quad\gamma=0.12\quad\delta=0.045\nonumber\\
\beta&=&0.58\quad\gamma=0.115\quad\delta=0.05
\end{eqnarray}
where the two sets of numbers correspond to different ways of
determining the experimental input.  Obviously the slope parameter
$\beta$ is in general agreement with experiment.  However, the
situation is more complex for the quadratic components.  In particular
the calculated values for $\gamma ,\,\delta$ are in good agreement with the
measurement of Gormley et al. (or Amsler et al.) but {\it not} with
that of Layter et al.  However, experimental uncertainties are
significant and a new high statistics determination of the spectral
shape is needed.

Additional information is available by studying the neutral decay mode
$\eta\rightarrow3\pi^0$, for which Bose symmetry determines
the decay amplitude to be
\begin{equation}
{\rm Amp}(\eta\rightarrow 3\pi^0)=N^{000}|1+\epsilon(x^2+y^2)_{\rm sym}|^2
\end{equation}
where
\begin{equation}
(x^2+y^2)_{\rm sym}={1\over 3}\sum_{i=1}^3(x_i^2+y_i^2)=2y_{\rm sym}^2
\end{equation}
Here the dispersive calculation by Kambor et al. predicts
\begin{equation}
\epsilon=-0.028,\quad {\rm or}\quad \epsilon=-0.014
\end{equation}
depending on how the experimental input is handled.  On the
experimental side the only measurement of the energy dependence until
recently had been of limited accuracy
\begin{eqnarray}
\mbox{Amsler et al.}\cite{exp4}: \epsilon&=&-0.044\pm 0.046\nonumber\\
\mbox{Baglin et al.}\cite{exp5}: \epsilon&=&-0.64\pm 0.74\nonumber\\
\mbox{Abele et al.}\cite{exp6}: \epsilon&=&-0.104\pm 0.04
\end{eqnarray}
which is consistent with (but with large experimental uncertainty) the
dispersive calculation.  However, recently a new result of
uncprecedented precision has been announced from the Crystal Ball 
group at BNL\cite{cb}
\begin{equation}
\epsilon=-0.062\pm 0.006\pm 0.004
\end{equation}

Clearly this number is significantly larger than expected from the
Khuri-Treiman calculation, suggesting that new dynamical input is
involved.  This is not unexpected.  Indeed the calculation of Kambor
et al. utilized the effective chiral Lagrangian at ${\cal O}(p^4)$ is
input.  Clearly from the agreement with experiment this is the main
effect, but one also expects contributions from pieces of the chiral
Lagrangian of ${\cal O}(p^6)$ such as
\begin{equation}
{\cal L}^{(6)}\sim{F_\pi^2\over \Lambda_\chi^2}{\rm tr}[(\chi
U^\dagger+U\chi^\dagger)D^\mu UD_\mu U^\dagger]{\rm tr}(D^\nu UD_\nu U^\dagger)
\end{equation}
where $\chi=2B_0m$, with $B_0$ being a constant and $m$ is the quark
mass matrix, $\Lambda_\chi\sim 4\pi F_\pi\sim 1$ GeV is the chiral
scale.  The coefficients of such terms are
unconstrained by the strictures of chiral invariance and are
experimentally undetermined at present, since they arise at two-loop
order.  Nevertheless, their presence can lead to peices of the decay
amplitude of the form
\begin{eqnarray}
A&\sim&{\cal A}_1k\cdot q_ca_a\cdot q_b+{\cal A}_2(k\cdot a_aq_b\cdot
q_c+k\cdot q_b q_a\cdot q_c)\nonumber\\
&\simeq&m_\eta^4\left({{\cal A}_1\over 18}+{{\cal A}_2\over
9}\right)\left[1-{Q_\eta^2\over m_\eta^2}(x^2+y^2)\right]\nonumber\\
&+&{1\over
12}m_\eta^4({\cal A}_1-{\cal A}_2)\left[{2Q_\eta\over
3m_\eta}y-{8\over 27}{Q_\eta^2\over m_\eta^2}(y^2-x^2)\right]
\end{eqnarray}
If such a dynamical component is present then its size should be set
by chiral scaling arguments
$${\cal A}_I\sim {m_d-m_u\over \Lambda_\chi^2F_\pi^2}$$
and an isospin relation
\begin{equation}
\gamma^{+-0}(dyn)+\delta^{+-0}(dyn)=\epsilon^{000}(dyn)
\end{equation}
must exist between the quadratic parameters for the charged and
neutral channels.  Here the symbol $dyn$ indicates the dynamical ({\it
i.e.}, non-rescattering component of the coefficient in question and is
found by subtracting the theoretical value obtained from the
Khuri-Trieman calculation from the experimental quantity.  In this way
we find, using the Gormley numbers for experimental input,
\begin{equation}
\gamma^{+-0}(dyn)=-0.025\pm 0.015\quad\delta^{+-0}(dyn)=-0.02\pm 0.02
\end{equation}
and
\begin{equation}
\epsilon^{000}(dyn)=-0.034\pm 0.007\quad{\rm or}\quad -0.048\pm 0.007
\end{equation}
depending on the dynamical input chosen.  The comparison
\begin{equation}
\gamma^{+-0}(dyn)+\delta^{+-0}(dyn)=-0.045\pm 0.03
\end{equation}
vs.
\begin{equation}
\epsilon^{000}(dyn)=-0.034\pm 0.007\quad{\rm or}\quad -0.045\pm 0.007
\end{equation}
is obviously satisfactory within errors but again cries out for a high
precision measurement of the $\eta\rightarrow \pi^+\pi^-\pi^0$
spectrum, such as would be possible using WASA.  

Of course, there is one additional test which we can use.  Since
according to isotopic spin invariance the $3\pi^0$ amplitude at the
center of the Dalitz plot must be a factor of three larger than the
corresponding $\pi^+\pi^-\pi^0$ number, the total decay rates should
differ by the factor 
\begin{equation}
{\Gamma^{(0)}(000)\over \Gamma^{(0)}(+-0)}={3^2\over 3!}=1.5
\end{equation}
When rescattering corrections are included, the prediction becomes
\begin{equation}
{\Gamma(000)\over \Gamma(+-0)}=\left\{\begin{array}{cc}
1.43& \mbox{one loop}\\
1.41\pm 0.03& \mbox{Khuri-Treiman}\end{array}\right.
\end{equation}
Both numbers are consistent with the value quoted by the Particle Data
Group\cite{pdg}
\begin{equation}
\left({\Gamma(000)\over \Gamma(+-0)}\right)^{{\rm exp}}=1.404\pm 0.034
\end{equation}
and are in good agreement with the recent Crystal Ball measurement\cite{exp3} 
\begin{equation}
\left({\Gamma(000)\over \Gamma(+-0)}\right)^{\rm exp}=1.44\pm 0.09\pm 0.01
\end{equation}

Clearly there is plenty of challenge in the three-pion sector for an
eta facility, but there is also interest in examining the remaining
radiative modes $\eta\rightarrow \pi^0\gamma\gamma , 3\pi\gamma$.

\subsection{\bf $\eta\rightarrow\pi^0\gamma\gamma$}

For the decay $\eta\rightarrow \pi^0\gamma\gamma$ chiral symmetry does not
play a important role, but vector dominance {\it does}.  This can be seen from
the feature that there exists no contribution at all to this process from
the tree level two-derivative Lagrangian.  Rather the lowest order 
chiral contribution
arises at one-loop level---$\eta\rightarrow 3\pi ,K\bar{K}\pi\rightarrow \pi^0
\gamma\gamma$.  However, the $3\pi$ intermediate intermediate state is
suppressed by the factor $m_d-m_u$ while the contribution from $K\bar{K}\pi$
is suppressed kinematically.  To see this, we define the general decay
amplitude
\begin{eqnarray}
& &{\rm Amp}(\eta\rightarrow\pi^0\gamma\gamma )=D(s,t,u)\left[\epsilon_1\cdot
\epsilon_2q_1\cdot q_2-\epsilon_1\cdot q_2\epsilon_2\cdot q_1\right]\nonumber\\
&-&E(s,t,u)\left[-\epsilon_1\cdot\epsilon_2p\cdot q_1p\cdot q_2-\epsilon_1
\cdot p\epsilon_2\cdot pq_1\cdot q_2\right.\nonumber\\
&+&\left.\epsilon_1\cdot q_2\epsilon_2\cdot p
p\cdot q_1+\epsilon_1\cdot p\epsilon_2\cdot q_1p\cdot q_2\right]
\end{eqnarray}
Then from the one-pion-loop contributions from ${\cal L}^{(2)}$ we find
\begin{equation}
 D_\pi(s,t,u)= {\sqrt{2}\alpha\over \pi}
T(s) F(s,m_\pi^2), \qquad E_\pi (s,t,u)=0
\end{equation}
where $T(s)$ is the lowest order $\eta\rightarrow 3\pi$ amplitude given in 
Eq. \ref{eq:qq} and
\begin{equation}
sF(s,m_\pi^2)=1+{4m_\pi^2\over s}\ln^2\left({\beta (s) +1\over 
\beta (s) -1}\right) \quad {\rm with} \quad \beta (s)=\sqrt{s-4m_\pi^2\over s}
\end{equation}
with a similar expression obtaining for the kaon loop contribution.  
Calculation
of the associated rate yields\cite{44}
\begin{equation}
\Gamma_{\rm loop}(\eta\rightarrow \pi^0\gamma\gamma )\approx
4\times 10^{-3}\,\,{\rm eV}\quad{\rm vs.}\quad \Gamma_{\rm exp}(\eta\rightarrow
\pi^0\gamma\gamma )=0.84\pm 0.18\,\,{\rm eV}.
\end{equation}
Thus the one-loop chiral contribution plays a very minor role.  
Consider, however, 
the vector dominance diagram shown in Figure 8, for which  
\begin{eqnarray}
D(s,t,u)&=&{2\sqrt{3}\over 9}g_{\omega\rho\pi}^2\left({2eF_\pi^2g\over m_V^2}
\right)^2\left[
{p\cdot q_2-m_\eta^2\over m_V^2-t}+{p\cdot q_1-m_\eta^2\over m_V^2-u}\right]\nonumber\\
&\times&\left({F_\pi\over F_8}\cos\theta -\sqrt{2}{F_\pi\over F_0\sin\theta}\right)
\nonumber\\
E(s,t,u)&=&-{2\sqrt{3}\over 9}g_{\omega\rho\pi}^2\left({2eF_\pi^2g\over m_V^2}
\right)\left[
{1\over m_V^2-t}+{1\over m_V^2-u}\right]\nonumber\\
&\times&\left({F_\pi\over F_8}\cos\theta 
-\sqrt{2}{F_\pi\over F_0}\sin\theta\right)
\end{eqnarray}
and yielding for the decay rate
\begin{equation}
\Gamma_{\rm VD}(\eta\rightarrow\pi^0\gamma\gamma )=0.31\,\,{\rm eV}
\end{equation}

\begin{figure}[htb]
\begin{center}
\epsfig{file=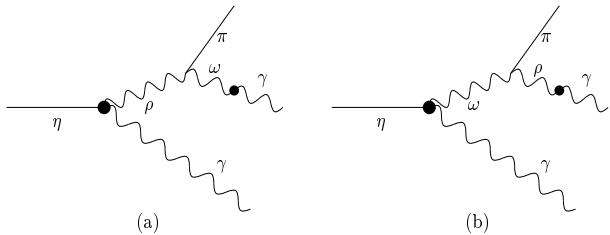,height=3cm,width=8cm}
\caption{Vector dominance diagram responsible for
$\eta^0\rightarrow\pi^0\gamma\gamma$.}
\end{center}
\end{figure}

Inclusion of other higher order effects such as the contribution from
a pair of anomalous 
terms---$\pi\pi\pi\gamma$ and $\eta\pi\pi\gamma$---coupled via a 
pion loop increases this estimate to about half the experimental
result, but a considerable discrepancy remains and should be the focus of
future experimental as well as theoretical work.  

A new number from the Brookhaven experiment was first
announced at this meeting by Nefkens
\begin{equation}
\Gamma^{exp}(\eta\rightarrow \pi\gamma\gamma)=(0.38\pm 0.11)\,\,eV
\end{equation}
is about a factor of two smaller than the Particle Data Group value.  
The problem with the previous measurements is probably associated with
eliminating the background from other much more probably neutral modes 
such as $\eta\rightarrow 3\pi^0$ and confirmation using WASA would
clearly be welcome.  In this regard, spectral shape measurements could
be helpful, although this will be difficult, since this is a low
branching ratio---$\sim 7\times 10^{-4}$---experiment.

\medskip
\subsection{\bf $\eta\rightarrow\pi\pi\pi\gamma$}
\medskip

The final mode which we shall mention in this report 
is $\eta\rightarrow 3\pi\gamma$, for 
which on the experimental side there exists at present only an upper 
bound\cite{pdg}
\begin{equation}
{\Gamma (\eta\rightarrow 3\pi\gamma )\over \Gamma (\eta\rightarrow 
\pi^+\pi^-\pi^0}|_{\rm exp} < 0.0024.
\end{equation}
Ordinarily the dominant component of a radiative mode such as this is due
to the the inner bremsstrahlung process, for which the matrix element is
\begin{eqnarray}
{\rm Amp}(\eta\rightarrow \pi^+\pi^-\pi^0\gamma )&\simeq&{\rm Amp}(\eta\rightarrow
\pi^+\pi^-\pi^0 )\nonumber\\
&\times& ie\epsilon^\mu\left[{(2p_++k)_\mu\over (p_++k)^2
-m_\pi^2}-{(2p_-+k)_\mu\over (p_-+k)^2-m_\pi^2}\right]
\end{eqnarray}
and it is experimentally difficult to distinguish any direct photon emission.  
However, an exception occurs when the non-radiative process is suppressed in
some fashion, such as occurs, {\it e.g.}, 
in the cases of $\pi^+\rightarrow e^+\nu_e\gamma$
and $K_L\rightarrow \pi^+\pi^-\gamma$, wherein the nonradiative process is 
small because of helicity suppression and CP violation
respectively\cite{ex}.  One might have anticipated the same enhancement 
mechanism to apply in our case since
the nonradiative reaction $\eta\rightarrow\pi^+\pi^-
\pi^0$ takes place only due to the relatively small u-d quark mass difference.
For example, the direct emission associated with the vector-dominance 
diagrams shown in Figure 9 could be expected to play an important role.
However, a careful analysis by D'Ambrosio et al. has shown that this 
unfortunately does not appear to be the case\cite{dam}. 
The contribution from
this direct---vector-dominated---mechanism is found to have the form
\begin{eqnarray}
A^\mu_{\rm direct}&(&\eta\rightarrow\pi^+\pi^-\pi^0\gamma )
={e64h_V\theta_V\over 3\sqrt{3}M_V^2F_\pi^4}\left[
p_\eta\cdot p_0g^\mu_{+-}+p_\eta\cdot p_+g^\mu_{-0}+p_\eta\cdot 
p_-g^\mu_{0+}\right]\nonumber\\
\quad
\end{eqnarray}
where
\begin{equation}
G_{ij}^\mu=k\cdot p_ip_j^\mu-k\cdot p_j p_i^\mu
\end{equation}
and the combination of Bose
symmetry and gauge invariance results in a suppression that makes 
such pieces even smaller than the pion loop component, which arise to
${\cal O}(p^4)$ and which are proportional to $m_u-m_d$.  In ref. \cite{dam}
it is estimated that the direct emission (DE) component is only a few percent
addition to the inner bremsstrahlung (IB), even at relatively large photon
energies
\begin{equation}
\left[(\Gamma_{\rm IB+DE}-\Gamma_{\rm IB})/\Gamma_{\rm IB}
\right]_{E_\gamma>90\,\,{\rm MeV}}\simeq 3.5\times 10^{-2} 
\end{equation}
which appears to be too small to make this a realistic experimental goal.

\begin{figure}[htb]
\begin{center}
\epsfig{file=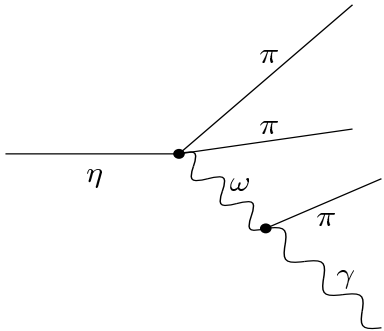,height=4.3cm,width=5cm}
\caption{Vector dominance diagrams responsible for
$\eta^0\rightarrow\pi^+\pi^-\pi^0\gamma$.}
\end{center}
\end{figure}

\section{Conclusions}

We have examined the "non-rare" decay modes of the eta meson---$\eta\rightarrow
\gamma\gamma , \pi^+\pi^-\gamma , 3\pi , \pi^0\gamma\gamma ,
3\pi\gamma$---in light of current theoretical knowledge and within the
general framework of chiral symmetry.  While there exist no
{\it striking} discrepancies observed with respect to any of 
these predictions, we
clearly identified problem areas wherein additional experimental
scrutiny, such as would be possible at CELSIUS, would
add significantly to our understanding.  These conclusions
can be summarized succinctly as follows:
\medskip

${\bf a:\,\, \eta\rightarrow \gamma\gamma}$---a resolution of the discrepancy between
the rates measured via the Primakoff effect\cite{22} and via QED 
production\cite{21} is essential to future progress in understanding the eta
system in general.

\medskip

${\bf b:\,\,\eta\rightarrow \pi^+\pi^-\gamma}$---a careful spectral
shape measurement would be useful in order check the spectral shape
with that predicted on fairly solid theoretical grounds from the
anomalous sector of QCD.

\medskip

${\bf c:}\,\,\eta\rightarrow 3\pi$---a precise spectral shape measurement 
is called for in order to determine whether the existing disagreement between experimental
findings and (what should be) solid theoretical predictions based on chiral
perturbation theory are cause by an inaccurate value for the d-u quark mass 
difference or are due to the importance of higher order final state interaction
effects.

\medskip

${\bf d: \,\,\eta\rightarrow \pi^0\gamma\gamma}$---a precision measurement of the
Dalitz plot distribution is suggested in order to learn the origin of the
existing disagreement between the experimental rate and that predicted from
vector dominance.

\medskip

${\bf e:}\,\, \eta\rightarrow 3\pi\gamma$---a determination of an 
actual rate instead
of the existing upper bound would be of interest, but theoretical
indications are that it will be difficult to detect other than the 
inner bremsstrahlung component.
\medskip

Clearly there is lots of interesting physics here and a marriage between
precise and solid new experimental data and careful and well-motivated 
theoretical analysis would, I predict, be a happy one, leading to the
offspring of a
new degree of understanding of an important component of low energy 
phenomenology.

\medskip

{\bf Acknowledgements}:  We thank John Donoghue for many clarifying 
discussions and the theory group at J\"{u}lich for their warm hospitality.
This work was supported in part by the Alexander von Humboldt Foundation and
by the National Science Foundation under grant PHY98-01875.


\begin{thebibliography}{99}

\bibitem{1} See, {\it e.g.}, M. Creutz, ``Quarks, Gluons and Lattices'' (Cambridge
University Press, Cambridge, 1983).
\bibitem{2} H.D. Politzer, Phys. Rep. {\bf 14}, 129 (1974).
\bibitem{3} S. Weinberg, Physica {\bf A96}, 327 (1979).
\bibitem{4} J. Gasser and H. Leutwyler, Ann. Phys. (NY) {\bf 150}, 142 (1984);
Nucl. Phys. {\bf B250}, 465 (1985).
\bibitem{5} J. Goldstone, Nuovo Cim. {\bf 19}, 154 (1961); J. Goldstone, A. Salam
and S. Weinberg, Phys. Rev. {\bf 127}, 965 (1961).
\bibitem{6} See, {\it e.g.}, S.B. Treiman in ``Current Algebras and Anomalies''
(Ed. by S.B. Treiman et al.) (Princeton Univ. Press, Princeton, 1985).
\bibitem{7} B.R. Holstein, Phys. Lett. {\bf B244}, 83 (1990).
\bibitem{8} See, {\it e.g.}, S. Gasiorowicz and D.A. Geffen, Rev. Mod. Phys. {\bf 41},
531 (1969).
\bibitem{9} S. Weinberg, Phys. Rev. Lett. {\bf 17}, 616 (1966).
\bibitem{10} E. Witten, Nucl. Phys. {\bf B223}, 422 (1983).
\bibitem{11} N.K. Pak and P. Rossi, Nucl. Phys. {\bf B250}, 594 (1985).
\bibitem{12} See, {\it e.g.}, B.R. Holstein, Int. J. Mod. Phys. {\bf A7}, 7873 (1992).
\bibitem{13} J.F. Donoghue and B.R. Holstein, Phys. Rev. {\bf D40}, 2378 and
3700 (1989).
\bibitem{14} B.R. Holstein, Comm. Nucl. Part. Phys. {\bf 19}, 221 (1990).
\bibitem{15} See, {\it e.g.}, D. Babusci et al., Phys. Lett. {\bf B277}, 158 (1992);
J.F. Donoghue and B.R. Holstein, UMass Preprint UMHEP-383 (1993).
\bibitem{16} See, {\it e.g.}, {\it PILAC Users Group Report on the Physics with
PILAC}, Los Alamos Report LA-UR-92-150.
\bibitem{17} See, {\it e.g.}, P. Herczeg, in ``Rare Decays of Light Mesons''
(ed. B. Mayer) (Editions Frontiers, 1991), p. 97.
\bibitem{gmo} M. Gell-Mann, CalTech Rept {\bf CTSL-20} (1961);
S. Okubo, Prog. Theo. Phys. {\bf 27}, 949 (1962).
\bibitem{18} J.F. Donoghue, B.R. Holstein and Y.-C.R. Lin, Phys. Rev. Lett.
{\bf 55}, 2766 (1985).
\bibitem{19} S.L. Adler and W.A. Bardeen, Phys. Rev. {\bf 182}, 1517 (1969).
\bibitem{20} Particle Data Group, Phys. Rev. {\bf D45} (1992).
\bibitem{21} S. Baru et al., Z. Phys. {\bf C48}, 581 (1990).
\bibitem{22} A. Browman et al., Phys. Rev. Lett. {\bf 32}, 1067 (1974).
\bibitem {fnt1} Here we use the $\eta\rightarrow
\gamma\gamma$ rate arising from QED production $e^+e^-\rightarrow 
e^+e^-\gamma^*\gamma^*\rightarrow e^+e^-\eta$\cite{21} rather than 
the value $0.324 \pm0.046\,\,{\rm keV}$ from the Primakoff 
effect\cite{22}.  It is clearly important to resolve this problem.
\bibitem{23} H. Behrend et al., Z. Phys. {\bf C49}, 401 (1991).
\bibitem{24} M. Gell-Mann et al., Phys. Rev. Lett. {\bf 8}, 261 (1962).
\bibitem{25} G. Ecker et al., Nucl. Phys. {\bf B321}, 311 (1989).
\bibitem{26} M. Bando et al., Prog. Theo. Phys. {\bf 73}, 1540 (1985);
T. Fujiwara et al., Prog. Theo. Phys. {\bf 73}, 926 (1985).
\bibitem{27} K. Kawarabayashi and M. Suzuki, Phys. Rev. Lett. {\bf 16},
255 (1966); Riazuddin and Fayyazuddin, Phys. Rev. {\bf 147}, 1071 (1966).
\bibitem{28} S.L. Adler et al., Phys. Rev. {\bf D4}, 3497 (1971); R. Aviv 
and A. Zee, Phys. Rev. {\bf D5}, 2372 (1971); M. Terent'ev, JETP Lett.
{\bf 14}, 94 (1971).
\bibitem{29} T.D. Cohen, Phys. Lett. {\bf B233}, 467 (1989); S. Rudaz, 
Phys. Lett. {\bf B145}, 281 (1984).
\bibitem{30} A. Bramon, E. Pallante and R. Petronzio, Phys. Lett. {\bf B271},
237 (1991).
\bibitem{31} Yu. M. Antipov et al, Phys. Rev. {\bf D36}, 21 (1987).
\bibitem{32} Ll. Ametller et al., Phys. Lett {\bf B276}, 185 (1992);
A. Bramon et al., ``The Da$\Phi$ne Physics Handbook'' (ed. L. Maiani,
G. Pancheri and N. Paver) (INFN, Frascati, 1992), p. 305.
\bibitem{33} P. Ko and T.N. Truong, Phys. Rev. {\bf 43}, R4 (1991).
\bibitem{akt} L. Ametller, M. Knecht, and P talavera, Phys. Rev. {\bf D64},
094009 (2001).
\bibitem{hol} B.R. Holstein, Phys. Rev. {\bf D53}, 4099 (1996).
\bibitem{tru} T. Truong, in ``Quantum Chromodynamics'' (ed. H.M. Fried
and B. Muller) (World Scientific, 
Singapore, 1999), p. 153. 
\bibitem{han} T. Hannah, Nucl. Phys. {\bf B593}, 577 (2001).
\bibitem{fnt2} For ease of calculation it
is sometimes useful to employ the simple analytic form
\begin{equation}
D_1(s)=1-{s\over m_\rho^2}-{s\over 96\pi^2F_\pi^2}\ln{m_\rho^2\over
m_\pi^2}-{m_\pi^2\over 24\pi^2F_\pi^2}F(s)
\end{equation}
\bibitem{34} M. Gormley et al., Phys. Rev. {\bf D2}, 501 (1970); 
J.G. Layter et al., Phys. Rev. {\bf D7}, 2565 (1973).
\bibitem{83} A. Abele et al., Phys. Lett. {\bf B402}, 195 (1997);
S.I. Bityukov et al., Z. Phys. {\bf C50}, 451 (1991).
\bibitem{baw} O. B\"{a}r and U.-J. Wiese, Nucl. Phys. {\bf B609}, 225 (2001).
\bibitem{fnt3} The transition can also occur due to electromagnetic 
effects but it is generally agreed that these are small.  See, {\it e.g.},
J.S. Bell and D.G. Sutherland, Nucl. Phys. {\bf B4}, 315
(1968); P. Dittner, P.H. Dondi, and S. Eliezer, Phys. Rev. {\bf D8},
2253 (1973); R. Baur, J. Kambor, and D. Wyler, Nucl. Phys. {\bf B460}, 
127 (1996).
\bibitem{os} H. Osborn and D.J. Wallace, Nucl. Phys. {\bf B20}, 23 (1970);
J.A. Cronin, Phys. Rev. {\bf 161}, 1483 (1967).
\bibitem{km} D. Kaplan and A. Manohar, Phys. Rev. Lett. {\bf 56}, 1994 (1986).
\bibitem{da} R. Dashen, Phys. Rev. {\bf 183}, 1245 (1969).
\bibitem{gle} J. Gasser and H. Leutwyler, Nucl. Phys. {\bf B250}, 539 (1985).
\bibitem{dhw} J.F. Donoghue, B.R. Holstein and D. Wyler, Phys. Rev. Lett.
{\bf 69}, 3444 (1992) and Phys. Rev. {\bf D47}, 2089 (1993); 
J. Bijnens, Phys. Lett. {\bf B306}, 343 (1993).
\bibitem{tr} T.N. Truong, Phys. Lett. {\bf B207}, 495 (1988).
\bibitem{ns} A. Neveu and J. Scherk, Ann. Phys. (NY) {\bf 57}, 39 (1970).
\bibitem{exp1} J. Layter et al., Phys. Rev. {\bf D7}, 2565 (1973).
\bibitem{exp2} M. Gormley et al., Phys. Rev. {\bf D2}, 501 (1970).
\bibitem{exp3} C. Amsler et al., Phys. Lett.{\bf B346}, 203 (1995).
\bibitem{al} A.V. Anisovich and H. Leutwyler, Phys. Lett. {\bf B375},
335 (1996).
\bibitem{kww} J. Kambor, C. Wiesendanger, and D. Wyler,
Nucl. Phys. {\bf B465}, 215 (1996).
\bibitem{cb} W.B. Tippens et al., Phys. Rev. Lett., {\bf 87}, 192001 (2001).
\bibitem{exp4} D. Alde et al., Z. Phys. {\bf C25}, 225 (1984).
\bibitem{exp5} C. Baglin et al., Nucl. Phys. {\bf B22}, 66 (1970).
\bibitem{exp6} A. Abele et al., Phys. Lett. {\bf B417}, 193 (1998).
\bibitem{43} J. Kambor et al., Phys. Rev. Lett. {\bf 68}, 1818 (1992).
\bibitem{pdg} D.E. Groom et al., Eur. Phys. J {\bf C15}, 1 (2001).
\bibitem{44} Ll. Ametller et al., Phys. Lett. {\bf B276}, 185 (1992).
\bibitem{ex} See, {\it e.g.}, G. D'Ambrosio and J. Portoles, 
Nucl. Phys. {\bf B533}, 523 (1998);  J.F. Donoghue and B.R. Holstein, 
Phys. Rev. {\bf D40}, 2378 (1989). 
\bibitem{dam} G. D'Ambrosio, G. Ecker, G. Isidori, and H. Neufeld, 
Phys. Lett. {\bf B466}, 337 (1999).
\end{thebibliography}
\end{document}